\documentclass[fleqn]{article}
\usepackage{graphics}
\begin{document}

\newcommand{\tab}{\hspace{5mm}}

\newcommand{\keywords}{blackbody radiation, Maxwell-Hertz electromagnetic theory,
detected signal} 

\newcommand{\PACS}{34.10.+x, 03.50.Kk, 84.47.+w, }

\title{On Temperature Radiation}
\author{S.\ L.\ Vesely$^{1}$, A.\ A.\ Vesely} 

\newcommand{\address}
  {$^{1}$I.T.B., C.N.R., via Fratelli Cervi 93, I-20090 Segrate(MI)
   \\ \hspace*{0.5mm} Italy \\ 
   }

\newcommand{\email}{\tt sara.vesely@itb.cnr.it, vesely@tana.it} 

\maketitle

{\small
\noindent \address
\par
\par
\noindent email: \email
}
\par

{\small
 \par
\par
\noindent{\bf PACS:} \PACS 
}

\begin{abstract}
In this work we compare two blackbody interpretations, those 
of G. Kirchhoff and M. Planck. We separate the problem of interpreting 
the blackbody spectrum from that of determining the mechanical 
equivalent of radiation. Then we propose that, if we set aside 
characterization of the blackbody, electromagnetism suffices 
to analyze spectral distribution and leads to interpreting it 
as determined by the receiver band-pass.
\end{abstract}

\section{Introduction}

Although the blackbody cannot be considered an elementary substance 
in a chemical sense, G. Kirchhoff considered it simple for thermodynamic 
interpretation of spectral properties. Indeed, he introduced 
it to demonstrate a theorem of capital importance in thermodynamics: 
that which fixes the relationship between emission $E$ and absorption 
$A$ of light and heat. According to the theorem, it makes sense 
to speak of thermal radiation if and only if for each pair of 
bodies $C_1$ and $C_2$ is equal to
 $E_1/A_1 = E_2/A_2$.
 The prerogative 
of the blackbody
$C_S$ is $A_S \equiv 1$. In the analysis of M. Planck 
however, the blackbody has lost the most fundamental property 
attributed classically to matter because it is an empty cavity 
at thermal equilibrium. It is clear that if the blackbody is 
identified with a ``macroscopic'' electromagnetic field, the latter 
appears as the most elementary of thermodynamic bodies, to wit, 
temperature radiation. But it is necessary to bear in mind that 
fields are the solutions of the Maxwell equations. As J. Clerk 
Maxwell explains, the equations describe the experiments of M. 
Faraday and give formal dress to his ideas. The physical interpretation 
of the solutions is less immediate. Therefore, identifying temperature 
radiation with an electromagnetic field neither specifies concretely 
the properties of Planck's blackbody nor interprets physically 
a solution of Maxwell's equations. Rather, that identification 
mixes two theoretical formulations having different assumptions 
to explain the phenomenon of heating by irradiation.

In this work we examine briefly the question of the existence 
of thermal emission on the basis of the properties which thermodynamics 
assigns it. On the one hand, electromagnetism does not specify 
any property for radiation while on the other it describes a 
mode of propagation incompatible with the second principle of 
thermodynamics. Indeed, it is linked historically with the development 
of telecommunications. But heat is not transferred from a warmer 
tank to a cooler tank by this technology but only a signal (the 
image) from a broadcaster to the receiving station regardless 
of the thermal gradients.

The most notable consequence of this analysis is the impossibility 
of determining an energy interpretation of the thermodynamic 
type  \cite{Becker 1964} for electromagnetism.

\section{A thermometer based on the brightness of hot bodies}

Around 1860 Gustav Kirchhoff wrote the famous article, ``\"{u}ber 
das Verh\"{a}ltnis zwischen dem Emissionsverm\"{o}gen und dem Absorptionsverm\"{o}gen 
der K\"{o}rper f\"{u}r W\"{a}rme und Licht'' in which he sought to 
give a definition of the thermometer alternative to that proposed 
by Lord Kelvin \cite{Thomson 1848}. In it he defined the ``blackbody''. 
Although this body, thus introduced, has a very well defined \textit{function} 
in thermodynamics -- that of measuring high temperatures -- it 
cannot have a structural \textit{connotation} quite as well defined 
for a reason which we shall try to clarify.

It is possible to use the luminous emission of a blackbody, for 
example the incandescence of the surface of a platinum wire, 
for measuring temperature if it is possible to calibrate in thermometric 
degrees a scale of luminous intensity or colors with its aid. 
If this is possible, the temperature can be read after the measuring 
blackbody has led to thermal equilibrium by intimate contact 
with the body whose temperature it is desired to measure. But 
light can be detected at a distance. Therefore it might be possible 
to measure the temperature at a distance by comparison of the 
light emitted by the body and that of a blackbody taken as standard 
(for instance by making use of a pyrometer). Since, unfortunately, 
different substances display appreciable variations in absolute 
emissive power, it can be asked whether the claim of graduating 
the temperature according to optical properties might have some 
basis.

Now let us clarify better the notion of optical thermometer. 
``Thermometer'' is any object suitable for reproducible temperature 
measurement. ``Optical'' means that it functions with no need 
of being placed in close contact with the body having unknown 
temperature. The other way round, any luminous object can function 
as an optical thermometer if it is possible to define \textit{the 
luminous equivalent of temperature}. In other words, \textit{in principle}, 
evaluation of temperature by the radiation emitted has sense 
if there is \textit{temperature radiation}. This, according to Kirchhoff, 
implies a value of the ratio between absolute emissive power 
and absorptive power dependent only on temperature. There are 
two alternatives.

A.\tab The first assumes that, experimentally, a sufficiently broad 
class of materials has well-defined thermal behavior and is linked 
to a radiative process. Logically, temperature is then conceived 
as an equivalence relation between the bodies. Since thermodynamics 
admits that a system can pass from the initial state to the final 
state only by traversing states of equilibrium, the thermodynamic 
parameters of the \textit{process} can be attributed to the \textit{state} 
of the system. If the radiative process passes through states 
characterized by temperature alone, then a function of the luminous 
emission measures it.

B.\tab The second alternative conjectures that all energy exchanges, 
and thus those of radiating energy too, are governed by thermodynamics. 
That is to say that the heat radiation exchanged is included 
in the energy balance and that radiation spreads from the hotter 
bodies to the colder ones. In the latter case a light parameter 
must be among the variables characterizing the process and must 
depend on temperature.

Alternative B requires that measurements depend on the initial \textit{temperature 
difference} between the thermometer and the object. Clearly, the 
same requirement applies to thermometers which function by contact. 
Except that it is known that the latter do not satisfy it. Specific 
heat $\chi = dQ/dT$ was introduced just because heating of 
bodies originally at different temperatures and placed in contact 
depends on their nature as well as their respective initial temperatures. 
Incidentally, the heat exchanged to reach the temperature of 
equilibrium, measured so as to be an additive magnitude, could 
replace temperature in thermodynamics. The problem is that it 
is not at all obvious what heat is. On the contrary, until now 
it is known only what it isn't; it is not a variable of state.

Let us explain this last statement better. Historically, Rumford 
was the first to interpret the need for continuous cooling when 
profiling cannon barrels in such a manner that the production 
of shavings produced heat. Hirn and Joule used industrial systems 
to appraise \textit{as accurately as possible} the amount of heat 
produced in relation to the work carried out by systems and agreed 
to interpret their measurement by saying that heat is, like mechanical 
work, a form of energy. The effort was considerable even if, 
the principle being established, the values found by those measurements 
can be checked in the laboratory with instruments which heat 
up little and convert little work into little heat, that is, 
when considerable fluctuations occur. But if the heat produced 
work in accordance with a constant relationship, the temperature 
might be associated with the work of expansion of a gas thermometer. 
That heat cannot be considered a variable of state when the system 
exchanges work was the unfortunate fact if we like.

After this short digression let us go back to temperature measurement 
and introduce the thermodynamic thermometer of Lord Kelvin with 
which we propose to compare Kirchhoff's. Lord Kelvin, as everybody 
knows, managed to define temperature independently of the particular 
thermometer basing himself on a cyclic process\footnote{This 
is not a \textit{periodic} process in the sense that it spontaneously 
takes on the initial conditions at the end of the period but 
one which again takes on the initial value of the internal energy 
($\Delta U = 0$) when the other variables of state are taken back to 
the initial values, i.e. $\Delta Q = \int pdv$ on the cycle. In other 
words, the process is not identified with rotation of the vanes 
of a steam turbine and it is not inverted by inverting the direction 
of rotation of the vanes.}.
Thanks to him, thermodynamic temperature 
is an additive magnitude. But it is also an absolute magnitude 
in the sense that the zero on the scale is not arbitrary. The 
choice, which expresses giving up defining the thermometer operationally, 
became necessary due to the Joule-Kelvin\footnote{ Called also 
``Joule-Thomson effect''. It is the effect which allows obtaining 
dry ice from carbon dioxide at surrounding temperature without 
doing work (for the work of expansion of freezing water, compare. 
Mach, p. 234\cite{Mach 1919}). As this means that the material states associated 
with thermodynamic processes are not unequivocally defined by 
the parameters, it leads to rejecting Alternative A.} 
effect. Lord Kelvin, after being persuaded that in any case an 
ill-defined thermometer does not allow establishing a unique 
mechanical equivalent of heat, redefined the thermometric scale 
on Carnot's ideal gas engine
\footnote{
Let us recall that in the 
Carnot cycle an ideal gas is, let us say, constrained to do work. 
The motor consists of one insulated cylinder with the exception 
of the base, and is equipped with one free thermally insulated 
piston. To work in the adiabatic/isothermal cycle it is placed 
alternately on an insulated stand or on one or the other of two 
stoves which are kept at temperatures $T_1$ and $T_2$ $K$ respectively. 
Its thermodynamic cycle takes place then between two temperatures 
$T_1$ and $T_2$ which are prevented from balancing themselves at 
the $T_{mean}$ such that $T_1 < T_{mean} < T_2$. Nor is the 
number of rpm mentioned because the cycle must function under 
conditions of thermodynamic equilibrium. This essential hypothesis 
excludes that a pressure difference be created between external 
and internal pressure at the cylinder (i.e. there must not be 
friction between the piston and the cylinder) and that the heat 
flows by conduction between the thermal baths and the piston. 
This hypothesis failing, Carnot's machine might not at all perform 
its duty despite the thermal gradient. Carnot's machine differently 
from Watt's or Otto's, which \textit{do not satisfy the hypothesis}, 
has the prerogative of being reversible. In this context, ``reversible'' 
means essentially that if a Carnot machine is connected to a 
Carnot refrigerator between the same two temperatures, it still 
must not be possible to transfer heat from the colder thermostat 
to the other. The postulate that the energy balance per cycle 
of the two (perfect) connected machines is null is equivalent 
to the postulate that the Carnot machine is ideal or reversible. 
It is also important to remember that the second principle, taken 
for the Carnot cycle, is universally valid in thermodynamics. 
Indeed, the energy statement is \textit{independent} of the particular 
mechanism considered.
}
. The reversibility of the cycle allows unequivocally defining 
the mechanical equivalent of heat. In addition, very general 
considerations on the impossibility of perpetuum mobile of the 
second kind allow keeping the same value of the equivalent, independently 
of the absolute temperature value. In this manner however there 
appears to be a thermodynamic state well defined at temperature 
zero. In addition, \textit{the choice of defining temperature on 
a rational basis\cite{Bryan 1907}instead of empirically, makes thermodynamics 
a prescriptive theory}.
In other words, this thermodynamics is 
not a discipline which describes facts subject to measurement 
error, but a mathematical one which deduces ideal universal processes 
from equations. Functional mathematical dependence \textit{defines} 
the thermodynamic potentials of energy, enthalpy, free energy, 
and entropy. Mathematics itself cannot be correct ``on average'' 
but only correct or, if the equations are incoherent, incorrect. 
This seems to us to be the manner in which a thermal phenomenon 
is measurable without thermodynamics being phenomenological.

The approach of Kirchhoff diverges from that of Lord Kelvin because 
the German physicist starts from the assumption that thermodynamics 
must be prescriptive. \textit{His blackbody is a thermometer because 
it satisfies the same relationship between heat and temperature 
that the absolute thermometer satisfies and not because it exists 
in our world.} This means that he does not consider the demonstration 
of his own theorem like a description of an observation.

In the demonstration, Kirchhoff takes into consideration any 
body $C$ having as its property that its radiation depends only 
on temperature and on wavelength since in any case $C$ can be colored. 
He shows that the process or absorption/emission is cyclical 
and is in agreement with the principles of classical thermodynamics 
(alternative B). For this purpose he makes use of a characteristic 
property of the mirror, which, in the case of the blackbody, 
is summarized, ``A mirror which reflects a blackbody is black''. 
Since, the other way around, ``an optically machined black glass 
is a mirror'', it is seen that at thermodynamic equilibrium the 
absorption/emission process is cyclical because it has the (optical) 
effect of a reflection. According to us, since heat does not 
produce work in accordance with a fixed relationship, the (mechanical) 
effect of a pressure cannot be attributed without further ado 
to optical reflection. Indeed, a ``Carnot cycle'' written to justify 
heat conduction between two thermal baths with different temperatures 
does not justify concomitantly the mechanical work of a thermal 
machine. In the same way, the substitute for the Carnot cycle 
proposed by Kirchhoff to justify the use of the pyrometer does 
not justify the mechanical work carried out by an electric motor. 
The $E/A$ relationship between emissive and absorptive power of 
the body $C$ measures temperature because the internal energy changes 
by $\Delta U$ due to the effect of heat $\Delta Q = \chi\cdot\Delta T$ 
absorbed with the change in temperature of the stove, where $\chi$ 
is precisely the specific heat of the body. Under the new conditions 
of equilibrium, $E/A$ measures the new temperature of the body 
$C$.

Instead of going into the details of the demonstration, let us 
consider the adherence of the terms to the facts. The first part 
of the thesis (stated in his num. 3), or the rule that emission 
be proportional to the absorption of a body for each wavelength 
of the radiation, is not in conflict with experience. Instead, 
the assertion that there is temperature radiation is not supported 
experimentally at least in three circumstances, which we recall 
below. Indeed, the theory postulates that at thermal equilibrium 
radiation depends only on temperature and possibly on wavelength, 
that is it takes on a \textit{homogeneous and isotropic distribution 
in space}\footnote{This does not occur ``on average'' but because 
the independence of radiation energy from spatial variables was 
postulated. In thermodynamics, wavelength is not in relation 
with the normal vibration modes of a box.}.
But, experimentally, 
the emission of electromagnetic radiators, called \textit{antennas}, 
does not possess the required characteristics of homogeneity 
and isotropy, nor does it take them on at thermodynamic equilibrium 
with a cylindrical, cubic or prismatic metallic container with 
optically machined walls. Not even the light of a common incandescent 
light bulb generally displays an efficient emission/absorption 
mechanism to establish thermodynamic equilibrium; if it remains 
lit and is visible from any point in a room it does not evolve 
towards a situation of ``equilibrium'' illumination characterized 
by the impossibility of distinguishing it from the wallpaper. 
In other situations involving measurement of power distribution 
it is not at all based on equilibrium; to affirm that radiation 
from the sun goes to thermal equilibrium with a thermometer located 
here on earth is ticklish from an experimental and a strictly 
theoretical viewpoint.

In conclusion, the operating principle of the optical pyrometer 
is justified by thermodynamics (alternative B) but the absorption/emission 
process is as ideal as the Carnot cycle.

\section{The quantum energetics interpretation}

In the preceding paragraph we saw that in thermodynamics the 
requisite of a thermometer, or indifferently an object whose 
temperature can be measured, is to possess a temperature independent 
of chemical composition or structure. We also saw that Kirchhoff 
assumed a universal principle, which governs energy exchanges 
(alternative B of the preceding num.) rather than attributing 
energetic states to matter (alternative A). Is it possible as 
an alternative to identify radiation with a form of energy?

M. Planck answered affirmatively. Indeed, he borrowed the statistical 
mechanical interpretation of entropy expressly for interpreting 
by a characteristic function the form of spectral radiation of 
hot bodies on the \textit{most updated} data\cite{Planck 1901}
\footnote{From approximately 
$1\mu$ up to $18\mu$, for temperatures from 87 $K$ to 
1650 $K$.}
of O. Lummer and E. Pringsheim. Now, in these experiments 
the oven is empty\footnote{
In this context, ``empty'' stands 
for ``containing a diathermanous medium'', eventually at a low 
pressure (in the theory of heat, diathermanous means transparent).}.
Hence the void exhibits special dependence on the wavelength 
of the emission of the blackbody better than any hot material 
object. That is to say, the property of being black belongs to 
radiation.

The characterizations of radiation which come out of the classical 
optics framework develop, in our opinion, along three mutually 
independent lines, to wit, statistical (*), energetic (**) and 
electromagnetic (***). Since Planck desired to underscore the 
originality of his approach as compared to that adopted by W. 
Wien or L. Boltzmann for extrapolating the \textit{less updated} data, 
he has become indeed the promoter of the explicative use of mechanical 
modeling (*) while continuing to insist that he wishes to pursue 
the work of clarification of the Maxwell equations (***) undertaken 
by H. Hertz (**)
\footnote{He expressed himself thus, ``the job 
of finding a modification [of the Maxwell theory] allowing for 
the new facts without sacrificing its best parts seems to me 
much more promising. In accordance with this criterion, the theory 
of radiation developed by myself will, in my opinion, reveal 
itself to be the most fruitful [of those thus far mentioned]. 
It is divided in two perfectly distinct parts, to wit, the electrodynamics 
theory of elementary oscillators and the statistical theory of 
these oscillators''. Ann. Phys. 31:758 (1910).}.
In short, Planck 
proposes to go beyond the energetic formulation of electromagnetism 
by making use of the atomic hypothesis of matter. We clarify 
below why, in our opinion, Planck's formula of irradiation from 
hot bodies does not represent energy distribution neither according 
to Boltzmann nor according to Hertz.

\subsection{States of light (*)}

The division between classical (''macroscopic'') 
system and surrounding environment (the rest of the world) is 
typical of the thermodynamics approach to the study of energy 
exchanges. One imagines the system being studied enclosed by 
a geometrical surface that ideally separates it from the environment. 
Exchanges of mechanical energy, heat, or even mass through the 
surface are considered measurable with instruments like the thermometer 
and the calorimeter. It is also thought that the system relaxes 
spontaneously until it settles into an indefinitely stable condition 
as a consequence of the failure of the constraints responsible 
for the initial differences in temperature, pressure and volume 
between system and environment. Classically, the thermodynamic 
potentials are attributed to the process and the accent is placed 
on the overall energy balance between the initial stable state 
and the final state of the system. But the definition of the 
absolute thermometric scale determines the problem of characterizing 
the states of matter in thermodynamics. The first theory which 
seeks to trace the thermal properties of the states back to dynamics, 
takes the name of kinetic theory. The latter does not consider 
the thermodynamic states as mere ``calculative expedients'' but 
associates with them dynamic properties deduced from an ad hoc 
model. Since dynamics is based on observations, it seems that, 
if a model allows justification of mathematical expressions of 
the state functions, thermodynamics will be refounded on physical 
principles.

According to Boltzmann, kinetic modeling is obtained by separating 
the thermodynamic system into a \textit{large number}
\footnote{ 
In a number of the order of that used by Avogadro during the 
volumetric determination of chemical reactions.}
of equal elements, 
atoms or molecules, individually satisfying the laws of classical 
mechanics. Indeed, for the purposes of \textit{dynamics}, any material 
system can be represented by marking out its elements (material 
points) subject to conservative forces. But each \textit{thermodynamic} 
system is supposed to be a material system and therefore the \textit{same} 
separation into elements (''microscopic'' subsystems) 
must apply. The correspondence between the variables which dynamics 
associates with the number of material points (considered as 
components of a ''macroscopic'' system) and the thermodynamic 
variables is not one to one. This is clear, according to Boltzmann. 
Indeed, thermodynamics labels only the states of equilibrium, 
that is the stable ones. And it must be possible to label them 
in an univocal manner, i.e. by adding a sufficient number of 
particular ''thermodynamic'' parameters, because 
the thermodynamic problem is that of univocal characterization 
of states. In a word, the physical properties of the system represented 
by thermodynamic functions do not apply to each atom but emerge 
from collective behavior at \textit{thermodynamic equilibrium}
\footnote{
The common indicator diagrams in the Cartesian plane connect 
only possible states of equilibrium. Therefore, on those diagrams 
there are no states different from those of equilibrium. That 
equilibrium is dynamic only for mechanical reduction usually, 
despite the mechanical conditions of equilibrium being assured 
by statics. Furthermore, in classical thermodynamic representation 
the system does not evolve from initially assigned conditions 
according to a position-time graph because neither positions 
nor time are part of the representation. So, formal replacement 
of the averages on the ''spatial'' variables with 
''temporal'' averages has no justification in thermodynamics.}. 
Since neither the laws governing the dynamic elements
\footnote{ 
Not those governing the evolution of the density of probability 
of the dynamic states.}
nor those which apply to the overall 
system are considered statistical in classical physics, what 
is statistical in the kinetic model is the collective behavior 
at equilibrium; it is governed by the law of large numbers.

An interpretative problem of thermodynamics emerges because the 
magnitudes involved in atoms, from which the system is composed 
according to the kinetic model, cannot be measured \textit{in principle} 
with classical measuring instruments. In other words, an average 
value is not the result of a statistical regularity with which 
a measurement error is associated experimentally. It follows 
that the law of large numbers is a mathematical principle. Roughly, 
the principle specifies the locution, \textit{``to converge in probability 
to a variable''}
\footnote{
We are discussing equilibrium states. 
Otherwise, dynamic time, defined in the dynamics of conservative 
systems basing on oscillatory or rectilinear motion, is not automatically 
the parameter governing thermodynamic relaxation also.
}
. This 
means that collective behavior is a probabilistic regularity 
in the specific sense that it assumes application of the theory 
of probability. This is the mathematical model on which the kinetic 
theory is based.

If the laws of thermodynamics must agree with the mathematics 
of the kinetic theory, their new interpretation differs from 
the classical one. Indeed, if the mathematical relation between 
the two types of variables - thermodynamic and mechanical - must 
respect the mathematical theory of probability, \textit{linear combinations} 
of dynamic variables enter into thermodynamic functions, all 
the possible elementary events must be known in advance, the 
sure event and the impossible event are in the field of events, 
et cetera
\footnote{
As thermodynamics is a physical theory and 
Boltzmann's atoms are microscopic, it is unclear which mechanical 
aspect determines the thermodynamic behavior of a system. But 
having reached this point it is possible to exemplify at least 
the type of relationship sought. It is a matter of connecting 
the roto-translation in the space of a tossed coin -- movement 
that can be filmed in the desired detail and reproduced by projection 
in ``as many identical examples as wanted'' -- with the quantitative 
evaluation of the possibility that a given side will appear on 
top.}
. What determines the classical probabilistic behavior 
of the system? Presumably the ''microscopic'' elements 
pertaining to a given thermodynamic state are shared equally 
as regards initial conditions. So the elementary events would 
be the sets of initial conditions. Since the number of events 
is very great but finite, the ``sharing process'' is not random. 
Indeed, the events can always be put in biunique correspondence 
with a finite subset of the set of natural numbers
\footnote{
To order numerable sets (for example, the elements of a set according 
to decreasing values) it is necessary to postulate an axiom equivalent 
to well-ordering.}
. Not even the real numbers associated with 
the temporal evolution $t$ at the dynamical level introduce any 
fortuitousness under the initial conditions. At the level of 
motion of one individual material point, to take the system back 
to the initial conditions at time $2t$ it suffices that the boundary 
conditions reverse the direction of motion on the trajectory 
at time $t$. More generally, it is possible to classify dynamic 
states according to symmetry properties and consider operations 
which reverse the states\cite{Lavis 2004}
\footnote{
To ''reflect'' 
a trajectory on itself, a correctly positioned reflecting surface 
is needed. To reflect n of them, n surfaces are needed. Mathematically, 
said conditions can not only be assigned but, without regard 
for complication, can be specified with the desired accuracy. 
Ignorance in principle of the experimental data associated with 
''microscopic'' subsystems does not imply any random measurement 
error on the data.}
. But it is necessary that the 
classification have a meaning. According to us, it has a meaning 
in mechanics but not in thermodynamics. Indeed, in hydrostatics 
the considerations on pressure direction rest on the fact that 
the surfaces on which hydrostatic pressure is exerted are precisely 
those of the container. But the separating surface between the 
thermodynamic system and the environment is largely arbitrary 
so that pressure in classic thermodynamics is defined in terms 
of elasticity of the (gaseous) system. In conclusion, evolution 
has no direct role in thermodynamics because the time defined 
in mechanics is not among the thermodynamic variables. However, 
the randomness at the origin of the irreversibility in Boltzmann's 
kinetic model does not depend on evolution but on the arbitrariness 
of the boundary conditions\cite{Tolman 1979}.

Furthermore, dynamic reduction presents two essentially mathematical 
difficulties. First, the system contains a finite number of elements 
so that all the average values calculated on it can be mapped 
as a function of a discrete variable. But thermodynamic functions 
are \textit{continuous} and in particular cover the whole real range 
of values between any pair of mean values calculated for the 
dynamic variables. That way, the dynamical states taken by one 
kinetic model only account for a set of measure zero in the graph 
of a given thermodynamic function. Quasi all thermodynamic states 
have no correspondent in the kinetic model. Second, the concept 
of \textit{mean value} pertaining to one element of the kinetic model 
is not well defined. When the values can be measured, we obtain 
a distribution of values from which mean values can be calculated 
according to a number of different criteria. On atoms, such measures 
cannot be taken in principle. The expected value must agree with 
the value calculated from thermodynamic laws or whatever method 
of obtaining averages will give rise to significant thermodynamic 
magnitudes? In the first case, i.e. if thermodynamics functions 
are expressions not only bound to the correct unit of measure 
but also mathematically \textit{assigned}, the statistical interpretation 
is misleading. As regards the second case, Einstein at last established 
the originality of the formula of spectral energy distribution 
of the blackbody with respect to the hypotheses of the kinetic 
theory of gasses. This means that a set of simple elements, let 
us say an ideal gas, is not a valid model of radiation. But Einstein's 
result really clarifies only that the model assumed by Boltzmann 
does not allow recovering the formula of Planck; it cannot show 
that the characteristic function which effectively interpolates 
the spectral form is an energy according to Boltzmann
\footnote{ 
On page 124 Einstein\cite{Einstein 1917} comments on the expression of the density 
of radiation.}
. Indeed, the mechanisms of emission/absorption 
assumed by Einstein have nothing in common with Boltzmann's collisions
\cite{Einstein 1907}. 
After all, classical statistical mechanics had never analyzed 
the consequences of inelastic collisions -- which give rise to 
emissions of the type of those which occur in radioactive phenomena 
-- on the state of a conservative mechanical system; nor did it 
admit directive shocks\cite{Einstein 1906}\cite{Born 1965}.

Lastly, if the blackbody has merely radiative consistency, the 
possibility of measuring thermal equilibrium experimentally with 
a conventional thermometer is lacking and therefore there is 
no longer any guarantee that the color or the intensity of the 
radiation measures temperature \textit{sooner or later} in the conventional 
sense.

\subsection{Extension of energetics to nonmechanical phenomena (**)}

The other way of saying that radiation is a form of energy is 
to formally include radiant energy in the energy balance (hypotheses 
B). While not taking support from dynamic modeling, the energy 
approach also interprets thermology with mechanics. To clarify, 
the cycle of a thermal machine in the usual Clapeyron diagram 
was explained as a balance even under the hypothesis that heat 
is a form of energy equivalent to mechanical energy. As underscored 
in the preceding num., a balance usually considers only exchanges 
and not absolute values. As to exchanges, the ''true'' 
Carnot cycle maximizes the transformation of heat into work. 
Despite this, the thermal difference on which the cycle functions 
does not yet allow univocally identifying the factor of conversion 
between heat and work because said factor also depends on the 
absolute temperature in degrees Kelvin. But it is possible to 
evaluate the energy output of heat in mechanical units. R. Clausius, 
who performed this calculation, adopted the subterfuge of introducing 
a new function of state instead of heat. This, being integrable, 
allows writing the energy balance between the initial and final 
states. Entropy without statistical significance is just that 
function of state which allows writing the heat absorbed per 
cycle by all thermal machines in mechanical units, i.e. in joules.
\cite{Planck 1900a}
Replacement of heat by entropy in balance calculation implies 
a conceptual abstraction even at the level of that which is imagined 
to be transferred between two bodies at different temperatures. 
As temperature increases, another mode of exchange of the heat 
substitute progressively appears, the mode called \textit{radiation}. 
This mode, just as heat conduction and its production by friction 
between mechanical parts in movement, \textit{is excluded from Carnot's 
cycle balance} (reversible)
\footnote{
If instead heat is assimilated 
with a fluid, the principle of conservation of heat is independent 
of the fact that heat produces work. Every mechanical interpretation 
of heat upstream of the dissipation problem would do.}.
For mechanical purposes it enters as a reduction of output. Despite 
the ``Kirchhoff cycle'' having null efficiency, he behaved as 
though Fermat's principle for optics allowed evaluating the mechanical 
equivalent of radiation. Even if we leave out of consideration 
the question of the unit of measure, it is not possible to calculate 
dissipated energy by leaning on thermodynamic states. As a matter 
of fact, this has been known for some time because the spectral 
lines of gaseous emission are typically associated with quantum 
jumps between given energy levels. But this is only a half solution. 
Indeed, it is clear that the representation of a transition cannot 
derive from the most meticulous study of the stability of the 
levels mentioned provided that the transition in question does 
not occur. As an example, if illumination whitens a coloring 
agent, the study of the chemical properties of the coloring agent 
before and after the reaction is different from that of the velocity 
of reaction as a function of illumination. But this \textit{is} half 
the solution. Indeed, we even judge as valid the interpretation 
of the associated emission as a mere signal. To be fair, and 
irrespective for a moment of the particular structural conjecture 
associated with the spectrum, all the chemistry based on spectral 
analysis identifies the substances by ''fingerprinting''.

Recapitulating, in classical thermodynamics, whether an absolute 
temperature is admitted or not, the hypothesis that the processes 
can pass through states all of equilibrium contains the designation 
of the thermodynamic states. Hertz himself admits at least that 
assigning an energy state in mechanics has sense. To give a meaning 
to ``non-mechanical'' energy, he calculates the balance 
between states in which slowly variable electrical magnitudes 
are balanced everywhere in ``physical space''
\footnote{
This 
approach, and hence its energy interpretation, fails if electrical 
measurements replace mechanical ones. Indeed, Hertz uses mechanical 
forces to balance unknown electrical quantities, not to establish 
once for all an electromechanical equivalent.}
by mechanical 
forces. If radiation occurs during the process, the radiant energy 
is not balanced but, Poynting's result for electromagnetism being 
valid, Hertz can hypothesize that radiation is transformed into 
heat by the Joule effect. Planck extends the balancing method 
to the dipole antenna (Hertzian oscillator), feigning that it 
would be possible to assimilate it with a mechanical oscillator.\cite{Planck 1900b} 
Planck's oscillators thus represent the mechanical counterpart 
which balances electromagnetic radiation (stationary). Admitted 
that it would be admissible to mediate (in the sample space) the 
energetic effect of electromagnetic oscillations on an oscillator, 
Planck considers that, at equilibrium, radiation is balanced 
on average. But an averaged oscillation is not at all periodic. 
Besides, conclusions about statistical populations of oscillators 
are not drawn in the space where bodies move, which for Hertz 
is precisely physical space. For lack of the mechanical balancer, 
Planck does not immediately represent an energy according to 
Hertz and therefore cannot assert that his formula represents 
a distribution of energy localized in the cavity of the oven 
and express it in joules. To be able to localize electromagnetic 
energy in space, irrespective of the velocity at which electrical 
magnitudes change, it is necessary to assume Einstein's space-time 
concept. Below, we are setting forth three points which in our 
opinion make the energy interpretation of the equations of electromagnetism 
problematic nevertheless.

Einstein's developments consist of this. Concomitantly with the 
principle of equivalence of mass and energy, Einstein postulated 
in 1905 that the expressions of the electromagnetic forces do 
not depend on the system of reference [used in mechanics for 
drawing position-time graphs]. In formulas, provided again 
that positive charges in motion generate neutral electrical current 
and putting the speed $c = 1$
\footnote{
Measurement is in Gauss 
units according to which $\sqrt{(\epsilon_0\mu_0)}=1$}
, he postulated 
that geometrical transformation of the frame from ${\bf E} = 0$, ${\bf H}$
magnetostatic to ${\bf E} = {\bf v}\times{\bf H}\to\partial{\bf H}/\partial t$
transforms an 
integral frame with a charge in slow motion into one integral 
with the magnet in which a force ${\bf F} = e({\bf v} \times {\bf H})$ 
acts on the charge (deflection of the electrical charge due to the 
effect of movement of the magnet at velocity ${\bf v}$). Instead, 
transformation from ${\bf H} = 0$, ${\bf E}$ electrostatic to
${\bf H} = -{\bf v}\times{\bf E}\to-\partial{\bf E}/\partial t$
transforms the frame integral with 
a magnetic needle into the one integral with the charge in motion 
in which the needle point is subject to the force ${\bf M} = - {\bf I} \times {\bf E}$
(deviation of the magnetic needle point due to the effect 
of a current of charges of velocity $-{\bf v}$). Because of the 
principle of equivalence, the field variables also have an energy 
interpretation for the microsystems and these correspond to geometric 
points or infinitesimal volumes. But before 1905 no one had ever 
asked that the diagrams of which thermodynamics makes use to 
represent energy processes enjoy some property of kinematic invariance. 
Indeed, the Galilean invariance principle established independence 
from position and velocity only for the laws of dynamics. Furthermore, 
a limit was set for the smallness of the systems to which thermodynamics 
is applied. Indeed, accurate determination of the mechanical 
equivalent of heat had required the availability of industrial 
means (Rumford, Hirn, Joule).

The first problematical point arises if we want to distinguish 
``physical space'' from the space-time diagrams of which classical 
mechanics makes use. Admitted that the relations indicated by 
Einstein represent Faraday`s electromagnetic induction and Maxwell's 
displacement current in Amp\`{e}re's equation respectively, the 
invariance of Maxwell's equation system involves mechanical transformation 
properties of the fields at small velocities. These transformation 
properties are by no means shared by ``physical space''. Again, 
the uniformity and isotropy prescribed for temperature radiation \textit{in 
the oven cavity} do not coincide with a superposition of stationary 
waves in space-time.

The second problem is that, because of the principle of equivalence, 
the position-time graphs of motion take on the same explicative value 
of the more usual (in thermodynamics) indicator diagrams. In 
a word, in the new representation the electromagnetic field energy 
variables depend on velocity in accordance with the principle 
of relativity (\textit{Einsteinian relativity}) without it being clear 
whether the arguments $x$, $y$, $z$ and $t$, of which they are in turn 
functions, are to be considered thermodynamic potentials or whether 
they are indices of position in bodies not uniformly heated as 
Fourier used them to describe the diffusion of the \textit{caloric}. 
Like Hertz, Fourier described in the ``physical space'' the mechanical 
aspect of a phenomenology while in thermodynamics usually the 
indices are thermodynamic parameters. This is not the place to 
discuss which domain it would be appropriate to assume for $x$, 
$y$, $z$ and $t$ in the Maxwell equations.

The third problematic point of Einstein's interpretation is the 
following. Until now we have identified radiative, electric et 
cetera phenomenology with the assigned mathematical expressions 
of the electromagnetic fields. Therefore, the expressions of 
the fields in empty space could already be considered equivalent 
to thermodynamic potentials. Instead, this is not quite correct 
because the equations with partial derivatives of which the fields 
are solutions establish only the dependence between variations 
and do not allow specifying functional expressions. But the fields 
in vacuo are obtained, for the mathematical theorem of existence 
and uniqueness of the solutions of differential equation systems, 
when the initial and boundary conditions are specified. Then 
the question is, how to obtain non trivial solutions. Indeed, 
assumed the relativity principle, and given the mathematical 
expression of an electromagnetic field, it is understandable 
that it does not reach thermodynamic equilibrium unless dissipation 
did not characterize that solution initially. In a word, for 
purely mathematical reasons, no solution of Maxwell's equations 
satisfying boundary and initial conditions, and not corresponding 
to ``blackbody radiation'' initially will merge with it at the 
end of all the transients.

In conclusion, the premise for interpreting electromagnetic fields 
as local energy perturbations in space and transport of electrical 
charges is that the arguments $x$, $y$, $z$ and $t$ take on a single 
meaning for all the theories involved and possibly that of coordinates 
of ``physical space''.

\subsection{Description of the electrical signal (received) (***)}

In this paragraph we shall again take up the clarification of 
Maxwell's equations undertaken by Hertz and got beyond by Einstein. 
Electromagnetism is a descriptive theory. Indeed, the system 
of the four partial differential equations stylizes as many key 
experiments of Faraday, to wit electrostatic induction, absence 
of magnetic monopoles, electromagnetic induction
\cite{Maxwell a}
and ``unipolar motor''. It doesn't seem to be a physics 
theory because Maxwell didn't clarify which of the solutions 
are elementary. Or better, he proposed the analogy with light 
transmission for the propagated solutions but without saying 
which characteristic of light he imagined represented by ${\bf E}$ 
or ${\bf H}$. Since the same theory includes the oldest conceptions 
of electrostatics and magnetostatics and it seems to make possible 
a unified treatment of them, the meanings attributed to the variables 
from the older theories were automatically accepted. Thus, experimental 
verification of the ''existence'' of Hertzian waves, instead 
of broadening the class of signal-solutions, favored the extension 
from electrostatics to electrodynamics of the mathematical magnitudes ${\bf E}$ 
and ${\bf H}$ (forces), and ${\bf D}$ and ${\bf B}$ (polarizations). Therefore, 
in writing the overall mechanical and thermal energy balance 
for those magnitudes, Hertz attacks the problem of determining \textit{the 
mechanical equivalent of the electromagnetic theory}
\cite{Hertz a}.

In the first place, he shows how, if desired, it is possible 
to take the equations of Maxwell from the potentials taken from 
the rival theory. His procedure consists of balancing with a 
system of electrical and magnetic forces the electrical and magnetic 
current variations that result because of mutual dependence on 
mathematical expressions.\cite{Bauer 2004} Obviously, he assumes that effective 
electrical magnitudes correspond to the mathematical corrections. 
He calculates by the induction principle (mathematical) the corrections 
necessary at each successive order. The procedure allows expanding 
in power series the forces and therefore the potentials from 
which they derive. The expansion in series of the potential function 
satisfies a wave equation and in addition allows taking the equations 
of electromagnetism in a quite immediate manner (as regularity 
conditions). But it is not possible to make reflections on the 
(mechanical) work carried out by the electrodynamics system because 
it is always balanced electrically.

In the second place he chooses as fundamental the experiment 
of Oersted, of the torsion of a compass needle near a battery-powered 
electric circuit, an effect analogous to that caused by a magnet. 
It ensues that the same mechanical effect can be caused either 
electrically or magnetically.
Therefore the formula: $(4\pi/c){\bf I}_0=\oint{{\bf H}d{\bf r}}$
is interpreted as though it linked 
together the measured value of the electric current and that 
of magnetomotive force\cite{Maxwell b}. For the behavior to be perfectly 
symmetrical, it would be necessary to observe an electrical effect 
of the magnets or even a mutual interaction between solenoids 
traveled by a direct current. In actual fact, Faraday observed 
that the abrupt movement of a conducting loop near a magnet produces 
an electric shock at the clamps of the circuit.
\footnote{
Maxwell 
num. 535. Vol. II. If the magnet M is replaced with a solenoid 
the variation in the movement of the conductor N also alters 
the current in the solenoid (M). It is clear, but we write it, 
that if the magnet can be replaced by a conductor, at the ends 
of which the current can be measured, the distinction between 
conductor-magnet M and conductor-circuit N is artificial, hence 
the case record is only apparent. In addition, Newton's third 
law plays no role. Indeed, as the symmetry between circuit N 
and magnet M is the consequence of varying only the current in 
one of the two conductors (M or N indifferently), induction takes 
place as the only effect. Faraday's induction is at the base 
of electromagnetic transmission and reception.}
In formula: $(1/c)\partial{\bf H}/\partial t = -{\bf curl}{\bf E}$.
 An effect that, according 
to Lenz, expresses an energy principle of homeostatic conception 
applied to Oersted's current.

Hertz's thermodynamic interpretation of electromagnetism depends 
on the hypothesis that the heat developed by the conductor is 
an \textit{integral part} of Oersted's observation\cite{Hertz b}
i.e. 
on the assumption that the electromagnetic magnitudes defined 
by the equations show some properties regardless of the mode 
of detection.
If the Faraday effect is the equivalent of Oersted's
\footnote{
Maxwell used Oersted's experiments to eliminate the magnetic 
hysteresis from all the linear treatment of electromagnetism. 
Consequently, those facts are contained in the equations. But 
they are not expressed in the so-called Amp\`{e}re-Maxwell equation. 
Indeed, the latter describes the experiment of Faraday's unipolar 
motor, termed later ``rotating field'' by G. Ferraris.}
experiment 
valid for variable fields, part of the electrical energy of the 
current must necessarily be dissipated in heat (in erg/second) \textit{even} 
in induction. In 1843 Joule had excluded that heat development 
was linked to induction; the latter appeared nevertheless linked 
to the resulting current increase. Thus it seemed that current 
flow always involves heat development. Today, on the contrary, 
we know that electric ''current'' in a solenoid is 
detectable with Oersted's needle under conditions in which heat 
is not developed
\footnote{
With superconduction the torsion 
of the magnetic needle is observed without accompanying heat. 
The process of ''creation'' of the magnetostatic field 
of the superconductor according to electromagnetism is in relation 
with the Faraday effect described by the integral of
$\partial {\bf H}/\partial t\div - {\bf curl}{\bf E}$
for the current which varies between 
closing of the circuit, $I = 0$ and stationary $I_0$ at which the 
potential difference at the ends of the solenoid vanishes
(then $L=\Phi/I_0$). In classical thermodynamics the process in which 
a material is transformed from (bad) conductor to superconductor 
is linked to the heat variation produced by the Oersted effect 
between when resistance $R = R_0$ and when $R = 0$. The reverse 
transformation, which does not pass through states of thermodynamic 
equilibrium, is called ''quench''.}
. Under these 
conditions we cannot associate with current energy (in mechanical 
units) dispersible in heat due to the Joule effect. But this 
mechanical contribution ${\bf jE}$ is what allows interpreting in 
energy terms all of the expression in the field variables, to 
wit:
$$ d/dt\bigl(1/8\pi\int\!\!\!\!\int\!\!\!\!\int({\bf E}^2+{\bf H}^2)dV\bigr)
+c/4\pi\int\!\!\!\!\int
({\bf E}\times{\bf H})\cdot{\bf n}dS=-\int\!\!\!\!\int\!\!\!\!\int{\bf jE}dV
$$
integrated on a volume $V$, on the boundary surface $S$ of which 
the electrical and magnetic forces do not vanish, without identifying 
the induced current with that obtained by Neumann
\footnote{ 
At nos. 542 and 543, vol. II, Maxwell explains the form of the 
theories of induction \textit{alternative} to his own. Hertz's work 
to which we allude is of 1884 vo1. 1 num. 17.}.
\cite{Jackson 1975}
\cite{Slater 1947}
\cite{Towne 1967}
Dropping the energetic interpretation of electromagnetic fields, 
there is no reason to think that the energy is saved or dissipated 
by induction. There is nothing counterintuitive in this interpretation; 
it is merely a matter of accepting that induction, as it is written, 
links two electrical variations without fixing any mode of detection.

In conclusion, mere identification of radiant heat with an entire 
theory (electromagnetism) not only does not allow writing the 
energy involved in the process of emission at equilibrium but 
even makes the subsequent development of thermodynamics incoherent. 
As to identification of black radiation with an observed phenomenon 
-- a problem totally outside of mathematics -- the first fine distinctions 
are surely prior to the theorem of Kirchhoff.

\section{Spectral distribution as the receiver's response}

In the preceding numbers we glanced at different formalizations 
of light phenomenology. Each of them answers certain requirements. 
In the past, it was not usual to make distinctions because it 
was not usual to think that mathematics is a language capable 
of representing even ideas in mutual disagreement. In addition, 
it was assumed that energy principles had a universal significance 
or it was even believed that they were automatically satisfied 
by any faithful description of nature. However that may be, ''verification'' 
of the predictions concerning radiant heat implied demands at 
the experimental level. One of these demands concerns the spectral 
characterization of radiation at equilibrium in the cavity of 
the oven (possibly electrically heated). The reason for this 
demand, strictly experimental, is to be sought in the fact that 
spectral analysis aimed at identifying chemical substances became 
popular at the same time. The principles of thermodynamics did 
not prescribe a particular spectral form for the black substance. 
Can they justify it when the experimenters find a credible one?

We propose the following two considerations.

i.\tab Not only the emissions of rarefied gasses apparently depend on 
the chemical constitution thereof but condensed matter also has 
characteristic emissions. Calling temperature radiation ''black'', 
practically all shades of ''gray'' are observed instead 
of it. Nor is it certain that the spectral form defined at the 
end of 1800 by the experimenters for the empty cavity of the 
oven at temperature T and attributed to black radiation is that 
of thermal equilibrium. Indeed we know that in the cavity of 
a device which selectively amplifies radiation the spectral composition 
of radiation at steady condition is different. Each of the experimental 
spectral forms is credible, but how many thermodynamic principles 
do we need to show the necessity and sufficiency of all?

ii.\tab Electromagnetism in turn does not explain the mechanism with 
which radiation of the sun takes place, or what sets off the 
lightning process, or what causes the luminescence of the firefly's 
glow. But the receiving and transmission modes according to Faraday 
and Hertz allow receiving them all and often attributing each 
of them to a specific transmitter. Thus, the theory after Faraday's 
experiments portrays as electromagnetic signals all of solar 
radiation, the flash and the firefly's glow indifferently.

Relying on these two considerations, we shall seek to explain 
why, according to us, the black radiation spectrum can be considered 
an antenna reception and can be interpreted as the receiver's 
band-pass.

First of all, we believe that it is necessary to clarify what 
was intended by \textit{temperature radiation} in the days of the 
crucial experiments. Says L. Graetz,\cite{Graetz 1906} ``It is common experience 
that a hot body in a cooler environment cools even leaving out 
of consideration conduction and convection. This cooling is called \textit{radiant 
heat.} Calorific radiation which leaves hotter bodies heats only 
the absorbent bodies but not the diathermanous ones of the environment. 
[\dots ] It is deduced [from the velocity of reception] that calorific 
radiation is of the same type as light radiation. [\dots ]''.

From the passage quoted it appears that thermal emission propagates 
differently from electromagnetic. Indeed, it spreads at great 
velocity from the hotter bodies to the colder ones. Now, since 
the diffusion process is not endowed with directivity, assuming 
that radiation spreads from hot to cold bodies, it is expected 
to be able to measure the absolute emissive power of a body only 
if the detector surrounds it completely and is kept at such a 
low temperature as to absorb entirely any radiation. As the thermometer 
cannot be taken to temperature $T = 0$ $K$ in the laboratory, the 
problem of measuring the spread of radiant heat has been faced 
by using a covering kept at a temperature other than zero but 
very well insulated. The radiation which no longer spreads due 
to the absence of the thermal gradients takes the name of \textit{temperature 
radiation}. First of all, it seems that an oven radiation is not 
a MASER because the radiation moves to equilibrium by diffusion 
instead of resonance, i.e. forming stationary electromagnetic 
waves\cite{carati2003}.
 Then, if there is temperature radiation, 
a pyrometer can measure the temperature at a distance. This means 
that the composition of a sample of radiation extracted from 
the oven, collimated, dispersed, analyzed and detected at the 
temperature of the laboratory is homogeneous with the spectral 
composition of the radiation at thermal equilibrium in the same 
empty oven. The external detector assigns to each frequency a 
characteristic intensity in accordance with a function $F(T,\nu)$. 
In the oven, which is empty in the specific sense that the air 
contained in the cavity does not heat by irradiation, being diathermanous, 
there is an internal detector. The internal detector, for example 
a bolometer, is set to measure the temperature of the oven cavity; 
for each value of the electric current measured there is a corresponding 
temperature. As the oven is empty, the internal detector attributes 
a single temperature to the radiation at thermal equilibrium. 
It cannot distinguish between their spectral components because 
the radiation is at thermal equilibrium and therefore all the 
frequency components are at the same temperature. But the external 
detector distinguishes the frequencies from each other because 
it allows assigning to each dispersed beam a different energy. 
Let us assume for convenience that the external detector is also 
a bolometer. The external bolometer measures a different electrical 
current for each frequency component of the black radiation impresses 
thereon (the laboratory air is also diathermanous). But the internal 
bolometer records a single current despite the radiative frequencies 
of the extracted sample being those of the radiation in the cavity. 
To us, it seems that the following circumstances are at the origin 
of the different behavior of the two bolometers. The body of 
the internal bolometer moves to thermal equilibrium with the 
internal walls of the oven by convection (in the meaning of the 
heat theory) despite the air being transparent to light. But 
the sensitivity of the resistance of the external bolometer depends 
on the particular striking spectral component in the first place 
because the radiation striking it was actually dispersed and, 
in the second place, because air at atmospheric pressure diffuses 
(in the optical sense) little at the so-called spectral windows.

The experimental verification, if we want to call it that, of 
the theorem of Kirchhoff in the case of black radiation seems 
to us unacceptable. Indeed, it would be a matter of taking the 
measured value of the emissive power for each wavelength back 
to the same instrumental sensitivity and so on for all the wavelengths 
actually irradiated by a cavity, however heated. But it is not 
necessary to know the emission band of the oven cavity to verify 
the theorem because this instruction is not in the terms. On 
the contrary, the theorem prescribes for the ideal blackbody 
the property of absorbing all the radiation regardless of temperature 
and wavelength, that is $A_S \equiv 1$ where $A_S$ is the absorption 
power of the blackbody. According to Kirchhoff, the emission 
of said blackbody must be proportional to absorption. Considering 
the fact that the property for the other bodies is given by a 
biratio, i.e. $E/A = E_S/A_S$, temperature being fixed, it can 
be put $E_S \equiv 1$ where $E_S$ is the emissive power of the blackbody. 
From the experimental verification of the theorem, with the correction 
for sensitivity, it might prove to be that not even the oven 
cavity is an ideal blackbody.

Rather than doing that, we are interested in pointing out that, 
whatever the manner in which the oven produces radiation and 
whatever be the homogeneousness of distribution of the intensity 
in the cavity, the technique of detection is the one used in 
telecommunications. That is, radiation with the characteristics 
of electromagnetic radiation propagation is detected. For this 
reason, allowance can be made for the sensitivity of the instruments 
by introducing the band-pass reception concept. Now we shall 
seek to make plausible that, given a body emitting with the mode 
hypothesized by Kirchhoff for the blackbody, the spectral form 
observed would be exactly the band-pass of the receiver. Indeed, 
let $E/A = \eta_\lambda \equiv F(T,\lambda)$ where $E$ and $A$ are the relative 
emissive and absorption power respectively of any body $C$ at temperature 
$T$ and wavelength $\lambda$.
Therefore, $E/A = 1\cdot\eta_\lambda(T,\lambda)$ where 
$\eta_\lambda$ or better $\eta_\nu(T,\nu)$ encodes the spectral response 
of the receiver. This is universal when the measuring instrument 
is assigned. But if the reference body is one of those which 
are usually called black without being black, i.e. if there isn't 
any temperature radiation, then the spectrum of the body obtained 
using a ``black detector'' would be
$E_G/A_G = \epsilon_G(T,\nu)$ while 
the spectrum actually recorded is the product of the irradiation 
of the body with the response in frequency of the receiver, i.e. 
$E/A = \epsilon_G(T,\nu)\cdot\eta_\nu(T,\nu)$ where $T$ is a frill because:

1.\tab the external detector is not at thermal equilibrium with the 
oven, and

2.\tab the radiation detected is not that of the blackbody.

\section{Summary and Conclusion}

We have sought in this work to expose why we think that formulating 
in classical thermodynamics the hypothesis that radiant heat 
is electromagnetic is of no use in accounting for either the 
distribution of energy of the blackbody or the irreversibility 
of the thermodynamic processes. As regards the first aspect, 
we have shown that temperature emission at thermal equilibrium 
does not depend by definition on the characteristics of the transmitter. 
But in the electromagnetic representation, irradiation depends 
on the initial and boundary conditions. As far as the second 
aspect is concerned, we have tried to clarify that thermodynamics 
and electromagnetism are different theories that respond to different 
requirements. It is possible to trace diagrams in classical thermodynamics 
but they do not depend on kinematic parameters, that is to say 
on spatial and temporal variables. But electromagnetism has been 
designed to enjoy the properties of invariance associated with 
movement in geometry. Although conduction, convection and radiation 
make the thermal machine cycle irreversible, the electromagnetic 
representation of transport phenomena does not account for irreversibility. 
Contrariwise, the emissive properties of the blackbody according 
to thermodynamics being assigned, analysis of the signal received 
allows concluding that the spectral form observed agrees with 
the receiver's band-pass.


\begin{thebibliography}{9}


\bibitem{Becker 1964} R. Becker, Fields and Interactions, Dover Publications Inc., New York (1982) (Ed. 1964 by F. Sauter)

\bibitem {Thomson 1848} W. Thomson, Opere di Kelvin by E. Bellone, UTET, Torino (1971) (Ed. 1848 by Phil. Soc. Proc. Cambridge)

\bibitem {Mach 1919} E. Mach, Die Prinzipien der W\"{a}rmelehre, 3. Auflage (1919), Minerva GMBH, Frankfurt/Main (1981)

\bibitem {Bryan 1907} G. H. Bryan, Thermodynamics, an introductory treatise dealing mainly with first principles and their direct applications, B. G. Teubners Lehrb\"{u}cher, Leipzig (1907)

\bibitem {Planck 1901} M. Planck, Ann. Phys. 4:553 (1901)

\bibitem {Lavis 2004} D. A. Lavis, http://arXiv.org/cond-mat/0401061 v.1 (2004)

\bibitem {Tolman 1979} R. Tolman, The Principles of Statistical Mechanics, Dover Publications Inc., New York (1979)

\bibitem {Einstein 1917} A. Einstein, Phys. Z. 18:121 (1917)

\bibitem {Planck 1900a} M. Planck, Ann. Phys. 1:669 (1900)

\bibitem {Planck 1900b} M. Planck, Ann. Phys. 1:719 (1900)

\bibitem {Einstein 1907} A. Einstein, Ann. Phys. 22:180 (1907)

\bibitem {Einstein 1906} A. Einstein, Ann. Phys. 20:627 (1906)

\bibitem {Born 1965} M. Born, Einstein's Theory of Relativity, Revised Ed., Dover Publications Inc., New York (1965)

\bibitem {Maxwell a} J. Clerk Maxwell, A treatise on Electricity \& Magnetism, III Ed. Dover Publications, Inc. (1954) New York Orig. Ed. 1891, vol. 2. nr. 530 and nr. 486

\bibitem {Hertz a} H. Hertz, Gesammelte Werke, vol. 1., p. 295, S\"{a}ndig Reprint Verlag, Vaduz Liechtenstein (1993), Ed. 1894

\bibitem {Bauer 2004} W. D. Bauer, http://arXiv.org/phys/0401151 v.1 (2004)

\bibitem {Maxwell b} J. Clerk Maxwell, A treatise on Electricity \& Magnetism, III Ed. Dover Publications, Inc. (1954) New York Orig. Ed. 1891vol. 2., nr. 481

\bibitem {Hertz b} H. Hertz, Gesammelte Werke, vol. 2. Untersuchungen \"{u}ber die Ausbreitung der elektrischen Kraft, Ed. 1894, S\"{a}ndig Reprint Verlag, Vaduz Liechtenstein (1993)

\bibitem {Jackson 1975} J. D. Jackson, Classical Electrodynamics, 2. Ed.,  nr. 6 John Wiley \& Sons, Inc., New York (1975)

\bibitem {Slater 1947} J. C. Slater and N. H. Franck, Electromagnetism, VIII, Dover Publications Inc., New York (1969), Ed. 1947

\bibitem {Towne 1967} D. H. Towne, Wave Phenomena, 6, Dover Publications Inc, New York (1988), Ed 1967

\bibitem  {Graetz 1906} Herausgeber A. Winkelmann, Handbuch der Physik, Band 3. W\"{a}rme, Johann Ambrosius Barth (1906)

\bibitem  {carati2003} A. Carati and L. Galgani, http://arXiv.org/physics/0312075 v.1 (2003)





\end{thebibliography}
\end{document}